\documentclass[12pt]{article}
\pdfoutput=1
\usepackage{comment,jheppub,xcolor,mathrsfs}
\usepackage{pgfplots}
\pgfplotsset{compat=1.17}
\usepackage{tikz}
\usepackage{feynmp-auto}
\usepackage[compat=1.1.0]{tikz-feynman}
\usepackage{lmodern}
\usetikzlibrary{decorations.pathmorphing}
\usepackage{appendix}
\tikzset{snake it/.style={decorate, decoration=snake}}

\author[a]{Mohsen Gheisarieha,}
\author[b,c]{Ramtin M.~Yazdi,}
\author[d]{Arash Arabi Ardehali$\,$}

\affiliation[a]{C.N. Yang Institute for Theoretical Physics, Stony Brook University,\\ Stony Brook, NY 11794, USA}

\affiliation[b]{Simons Center for Geometry and Physics,
Stony Brook University,\\ Stony Brook, NY 11794-3636, USA}

\affiliation[c]{Department of Physics and Astronomy, Stony Brook University,\\ Stony Brook, NY 11794, USA}

\affiliation[d]{Physics Department, Sharif University of Technology,\\
Azadi Avenue, Tehran, Iran}

\emailAdd{mohsen.gheisarieha@stonybrook.edu}\emailAdd{ramtin.mohasselyazdi@stonybrook.edu}\emailAdd{a.a.ardehali@gmail.com}

\title{
Integrable Spherical Brane Model at Large $N$}

\dedicated{Dedicated with gratitude to Sasha~Zamolodchikov on his 73rd birthday}

\begin{document}

\abstract{We study one of the simplest integrable two-dimensional quantum field theories with a boundary: $N$ free non-compact scalars in the bulk, constrained non-linearly on the boundary to lie on an
$(N-1)$-sphere of radius $1/\sqrt{g}$. The $N=1$ case reduces to the single-channel Kondo problem, for $N=2$ the model describes dissipative Coulomb charging in quantum dots, and larger $N$ is analogous to higher-spin impurity or multi-channel scenarios. Adding a boundary magnetic field---a linear boundary coupling to the scalars---enriches the model’s structure while preserving integrability. Lukyanov and Zamolodchikov (2004) conjectured an expansion for the boundary free energy on the infinite half-cylinder in powers of the magnetic field. Using large-$N$ saddle-point techniques, we confirm their conjecture to next-to-leading order in $1/N$. Renormalization of the subleading solution turns out to be highly instructive, and we connect it to the RG running of $g$ studied by Giombi and Khanchandani (2020).}

\maketitle

\section{Introduction}\label{sec:intro}

Dissipative quantum mechanics (DQM) has long served as a bridge between condensed matter and high-energy physics. A paradigmatic framework is the Caldeira–Leggett model of a quantum particle coupled to a bath of oscillators, which captures quantum Brownian motion and tunneling with dissipation~\cite{Caldeira:1981rx,Caldeira:1983jx,Caldeira:1983hj}. This model (and its two-level variant, the spin-boson model \cite{leggett1987dynamics,Weiss:2021uhm}) has provided deep insights into quantum impurity dynamics in contexts ranging from superconducting devices to the Kondo effect. In parallel, similar concepts arise in string theory: notably, Callan and Thorlacius showed that certain open string vacuum configurations can be mapped to Caldeira–Leggett-type dissipative systems~\cite{Callan:1990sf}. In their work, an open string with appropriate boundary conditions was found to behave like DQM at criticality, inheriting full reparametrization invariance from the string’s worldsheet conformal symmetry. These developments established a fruitful correspondence between open string theory and quantum dissipation, motivating further study of ``open'' quantum systems using field-theoretic tools.

In this paper, we explore a particularly elegant integrable model that lies at the intersection of DQM and string theory---often referred to as the \emph{spherical brane model}. It consists of $N$ free two-dimensional scalar fields $X^i(\sigma,\tau)$ (with $i=1,\dots,N$) with a nonlinear boundary condition that constrains the fields at $\sigma=0$ to lie on an $(N\!-\!1)$-sphere. Equivalently, the boundary degrees of freedom can be parametrized by an $N$-component vector field $\mathbf{n}(\tau)$ with $\mathbf{n}^2=1/g_0$, where $g_0$ can be thought of as a coupling. The boundary may be thought of as an open string endpoint constrained to move on a target-space sphere. The case $N=1$ (an $(N\!-\!1)=S^0$ boundary) corresponds to a two-state impurity and is related to the single-channel Kondo problem \cite{Lukyanov:2003rt}, while larger $N$ is analogous to multi-channel or higher-spin impurity scenarios. The connection with DQM arises by associating $\mathbf{n}(\tau)$ to the system, $\boldsymbol{X}(\sigma,\tau)$ to the bath, and the non-linear boundary condition to their interaction (a Lagrange multiplier implementation of the boundary condition makes the system-bath interaction more transparent). As we shall see in Section~\ref{sec:leading} (Eq.~\eqref{eq:Seff_YB}), integrating out the bulk fluctuations generates an effective action for $\mathbf{n}(\tau)$ whose nonlocal (in time) kinetic term encodes dissipation (as seen through its derivative expansion, for instance). 

A central application of this kind of model (with $N=2$) is the study of Coulomb blockade phenomena in quantum dots \cite{Averin:1986,feigelman2002weak,beloborodov2003two,Lukyanov:2003rt}. In these semiconductor nanocrystals, charging at low temperatures is impeded by Coulomb repulsion from electrons already occupying the dot. Although the blockade may be viewed naively as a simple energy constraint, dissipation qualitatively changes the physics: a tunneling event excites the electronic environment, leading to an orthogonality-catastrophe–type effect that makes the problem genuinely many-body dynamical \cite{Matveev1995,Matveev1996,Borda2006,Goldstein2010}. In the effective field theory description, $g_0$ quantifies the dimensionless resistance of the tunneling contact, while the nonlocal kinetic term for $\mathbf{n}(\tau)$ encodes the dissipative dynamics that govern electron transport and determine the scaling behavior of the blockade.

From a pure field theory perspective, the model serves as a prototypical example of a \emph{boundary renormalization group (RG) flow}. A Lagrange multiplier implementation of the boundary condition replaces the nonlinear constraint with a nontrivial interaction localized at the boundary, which generates an RG flow to Dirichlet boundary condition in the infrared (cf.~\cite{Kosterlitz:1976zz,Giombi:2019enr}).


The rich structure of the spherical brane model was explored by Lukyanov and Zamolodchikov who identified it as an \emph{integrable} boundary field theory \cite{Lukyanov:2003rt}. They investigated the model's response to a boundary magnetic field, $\boldsymbol{H}$, which couples linearly to the scalars at $\sigma=0$ and preserves the model's integrability. Based on the properties of the conserved charges expected to form the model's integrable hierarchy, they conjectured an all-orders expansion for the specific boundary free energy $\mathcal{E}$ (on an infinite half-cylinder in the large-radius limit) in powers of $\boldsymbol{H}$. This conjecture provides an explicit characterization of the theory's underlying algebraic structure.\footnote{The paper \cite{Lukyanov:2003rt} also considers in great detail the ODE/IM correspondence \cite{Dorey:1998pt,Bazhanov:1998wj,Dorey:2007zx} for small values of $N$, especially $N=2$. That aspect of the model is beyond the scope of the present work.}

The principal aim of the present paper is to perform a highly non-trivial check of the Lukyanov-Zamolodchikov conjecture for $\mathcal{E}$
using large-$N$ techniques. By treating $1/N$ as a small parameter, we employ the saddle-point method to compute the boundary free energy to next-to-leading order in $1/N$ and to all orders in the magnetic field $\boldsymbol{H}$. Our analysis begins with implementing the non-linear boundary condition through a Lagrange multiplier field $U(\tau)$. The leading-order solution then arises from a saddle-point evaluation of $\mathcal{E}$, while the subleading solution requires a one-loop calculation involving fluctuations of $U$. Renormalization of the UV divergences arising from the loop diagram turns out to be highly instructive. We hence dedicate three subsections to various aspects of it, making contact with path-integral measure normalizations in quantum mechanics and the RG running of $g_0$ as studied in \cite{Giombi:2019enr}.

Our results are in perfect agreement with the expansion proposed by Lukyanov and Zamolodchikov. This confirmation provides strong evidence that the integrable hierarchy of the model was correctly identified in \cite{Lukyanov:2003rt}. It also validates the power of the $1/N$ expansion in capturing the non-perturbative dynamics of this integrable boundary field theory, and motivates further explorations of the intersection of integrability, string theory, and the fundamental principles of dissipative quantum mechanics within this model (and those similar to it \cite{Zamolodchikov2006}) through large-$N$ techniques.

This paper is organized as follows. Section~\ref{sec:model} introduces the model and reviews the Lukyanov-Zamolodchikov conjecture for the series expansion of its specific boundary free energy $\mathcal{E}$ (on an infinite half-cylinder in the large-radius limit) in powers of $\boldsymbol{H}$. Section~\ref{sec:leading} presents the Lagrange multiplier reformulation of the problem and solves it at the leading order in the $1/N$ expansion by evaluating $\mathcal{E}$ on the large-$N$ saddle point. Section~\ref{sec:subleading} takes the computation of $\mathcal{E}$ to next-to-leading order, in particular by evaluating and renormalizing a one-loop diagram involving the Lagrange multiplier field. Section~\ref{sec:future} describes ways of improving our treatment, as well as extending it to more conceptual territories. Technical details of the large-radius asymptotic evaluation of the subleading solution are relegated to the appendix.

\section{The model and the Lukyanov-Zamolodchikov conjecture}\label{sec:model}

The theory of our interest \cite{Lukyanov:2003rt} consists of $N$ two-dimensional real non-compact bosons, denoted collectively by $\boldsymbol{X}$, on a cylinder parametrized by $0\leq\tau\leq2\pi R$ and $-\ell\leq\sigma\leq0$. We are interested in the limit $\ell\to\infty$. The semi-infinite cylinder can be mapped to a disk using the transformation $\frac{z}{R} = e^{(\sigma + i\tau)/R}$, where \( \sigma \in (-\infty, 0] \) and \( \tau \sim \tau + 2\pi R \).

We subject $\boldsymbol{X}$ to the Neumann boundary condition at $\sigma=-\ell,$ and to the constraint
\begin{equation}
    \boldsymbol{X}^2\big|_{\sigma=0}=\frac{1}{g_0},\label{eq:constraint}
\end{equation}
at $\sigma=0.$ These are compatible with the following action, which contains also a magnetic field $\boldsymbol{H}$ on the $\sigma=0$ boundary
\begin{equation}
    \mathcal{A}_\text{cyl}=\frac{1}{4\pi}\int_{-\ell}^0\mathrm{d}\sigma\int_0^{2\pi R}\mathrm{d}\tau\, (\partial_\sigma\boldsymbol{X}\cdot\partial_\sigma\boldsymbol{X}+\partial_\tau\boldsymbol{X}\cdot\partial_\tau\boldsymbol{X})-2\int_0^{2\pi R}\mathrm{d}\tau\, \boldsymbol{H}\cdot \boldsymbol{X}_{\!\sigma=0}(\tau).\label{eq:actionX}
\end{equation}
To see the compatibility note that $\delta_X\mathcal{A}_\text{cyl}=\frac{1}{2\pi}\oint\mathrm{d}\tau\, (\partial_\sigma \boldsymbol X-4\pi \boldsymbol H)\cdot \delta \boldsymbol X,$ neglecting the bulk equation of motion term, as well as the contribution from the boundary at $\sigma=-\ell$ due to the Neumann condition there. So the variational problem is well-defined if in addition to \eqref{eq:constraint} we impose the \emph{modified Neumann boundary condition}
\begin{equation}
    (\partial_\sigma \boldsymbol X-4\pi \boldsymbol H)\cdot\tilde{\delta}\boldsymbol X\,\big|_\partial=0,\label{eq:modNeuX}
\end{equation} with the tilde denoting restriction to the space of boundary functions satisfying \eqref{eq:constraint}.\footnote{As an example, for $N=2,$ writing $X_1=\frac{X_r}{\sqrt{g_0}}\cos\Theta$, $X_2=\frac{X_r}{\sqrt{g_0}}\sin\Theta$, and $H_1=H\cos\alpha$, $H_2=H\sin\alpha$, we have $\tilde{\delta}X_1|_\partial=-\sin\Theta\,\delta\Theta,\ \tilde{\delta}X_2|_\partial=\cos\Theta\,\delta\Theta$, and consequently the modified Neumann condition \eqref{eq:modNeuX} in terms of $\Theta$ becomes 
\begin{equation*}
    (\partial_\sigma \Theta-4\pi\sqrt{g_0}H\sin(\alpha-\Theta))\big|_{\sigma=0}=0.
\end{equation*}}

The theory is asymptotically free, in the sense that $g_0$ runs to zero as the UV cutoff $\Lambda$ goes to infinity \cite{Lukyanov:2003rt,Giombi:2019enr}. Alternatively, it develops a physical energy scale $E^\ast$, such that at distances $\gg \frac{1}{E^\ast}$ from the $\sigma=0$ boundary the effect of the boundary is that of the fixed (Dirichlet) boundary condition $\boldsymbol{X}_{\!\sigma=0} = 0$.

The observable of our interest is the partition function (see Eq.~(8) of \cite{Lukyanov:2003rt})
\begin{equation}
    Z=\lim_{\ell\to\infty}\ N\, e^{\frac{N\ell}{12R}}\, \int\mathcal{D}\boldsymbol{X}\, e^{-\mathcal{A}_{\text{cyl}}}.\label{eq:ZRH_def}
\end{equation}
See \cite{Lukyanov:2003rt} for explanation of the prefactor.\footnote{It primarily ensures that the $R\to\infty$ behavior of $Z$ is $\sim e^{-2\pi R\mathcal{E}}$ with coefficient $1.$}

We define the specific boundary free energy as
\begin{equation}
    \mathcal{E}:= -\frac{1}{2\pi R}\log Z\,.
\end{equation}
It was argued in \cite{Lukyanov:2003rt} that one can express $\mathcal{E}$ in the $R \rightarrow \infty$ limit as a power series in $\left( \frac{H}{E^\ast} \right)^2$ of the form:
\begin{equation}
    \mathcal{E}(H,E^*) = E^* \sum^{\infty}_{k=0} C_k \Big(\frac{H}{E^*}\Big)^{2k},
    \label{eq:sasha_free_energy_series}
\end{equation}
with $H^2:=\boldsymbol{H}\cdot\boldsymbol{H}.$ To remove the scheme ambiguity of the physical scale $E^\ast$, we use the convenient normalization condition of \cite{Lukyanov:2003rt} relating it to the zero-temperature susceptibility:
\begin{equation}
    \frac{1}{E^\ast} = -\frac{1}{2} \frac{\partial^2 \mathcal{E}}{\partial H^2}\bigg|^{}_{H=0}\,.
    \label{eq:norm_cond_intro}
\end{equation}

Taking advantage of the integrability of the model, and based on the expected forms of the integrals of motion within the framework of integrable boundary QFT (see e.g. \cite{Zamolodchikov2006}), it was conjectured in \cite{Lukyanov:2003rt} that the coefficients $C_{k\ge1}$ in \eqref{eq:sasha_free_energy_series} take the form
\begin{equation}
    C^{}_{k\ge1} = (-1)^k (2\Delta)^{1-k} \frac{\Gamma\Bigl((2k-1)\Delta\Bigr)}{k! \; \Gamma^{2k-1}(\Delta)} (2k-1)^{k(1-2\Delta)+\Delta-1}\,,
    \label{eq:sasha_full_coefs}
\end{equation}
where $\Delta: = \frac{1}{N-1}$.

The principal aim of the present paper is to verify the Lukyanov-Zamolodchikov conjecture \eqref{eq:sasha_full_coefs} to next-to-leading order in the $1/N$ expansion. Taking $\frac{H^2}{N\,E^{*2}}$ to be $\mathcal O(1)$ as $N\to\infty$, we extract the leading and next-to-leading parts of \eqref{eq:sasha_free_energy_series}:
\begin{equation}
\begin{split}
    \mathcal{E} &= E^* \; \sum_{k=0}^\infty { \frac{(1-2k)^{k-2}}{2^{k-1}k!} N\left(\frac{H^2}{N\,E^{*2}} \right)^{k} } \\
    &+\; E^* \; \sum_{k=0}^\infty {\left[k-1-(2k-1) \; \log(2k-1) \right]\frac{(1-2k)^{k-2}}{2^{k-1}k!} \left(\frac{H^2}{N\,E^{*2}} \right)^{k}},
\end{split}
\end{equation}
up to $\mathcal{O}(1/N)$ corrections. The first line is the $\mathcal O(N)$ leading order piece, and the second line is the $\mathcal O(1)$ next-to-leading order piece in the $1/N$ expansion.

\section{Leading order solution in the $1/N$ expansion}\label{sec:leading}

We incorporate the boundary condition \eqref{eq:constraint} into the action \eqref{eq:actionX} using a Lagrange multiplier term 
\begin{equation}
    +\frac{1}{4\pi} \int_0^{2\pi R}\mathrm{d}\tau\ 
 U(\tau) \Big(\boldsymbol{X}^{2}\big|^{}_{\sigma=0} - \frac{1}{g_0}\Big)\,,
\end{equation}
We also write $U(\tau) = E + u(\tau)$, with $E$ the expectation value of $U(\tau)$. To remove the linear dependence of the modified action on $\boldsymbol{X}_{\sigma=0}$ (the last term in \eqref{eq:actionX}), we define a shifted variable 
\begin{equation}
\boldsymbol{Y}:=\boldsymbol{X}-\frac{4\pi \boldsymbol{H}}{E}.\label{eq:XtoY}
\end{equation}
The action can now be written as
\begin{equation}
    \begin{split}
    \mathcal{A}_\text{cyl}[\boldsymbol{Y},u]&=-\frac{1}{4\pi}\int_\text{bulk}\boldsymbol{Y}\cdot\Box\boldsymbol{Y}-\frac{1}{4\pi}\int_\partial \big( \frac{(4\pi H)^2}{E}+\frac{E}{g_0}\big)\\
    &\quad+\frac{1}{4\pi}\int_\partial \big(\boldsymbol{Y}\cdot\partial_\sigma\boldsymbol{Y}+E\boldsymbol{Y}^2+u(\tau)\big[\big(\boldsymbol{Y}+\frac{4\pi \boldsymbol{H}}{E}\big)^2-\frac{1}{g_0}\big]\big).\label{eq:actionFinalY}
    \end{split}
\end{equation}
In this formulation, for the variational problem to be well-defined we need to impose the \emph{modified Neumann} (or more precisely \emph{Robin}) \emph{boundary condition}\footnote{This arises starting from the action in the form with bulk kinetic term $\partial Y\partial Y$ (rather than $Y\Box Y$) and then demanding $\delta_Y\mathcal{A}_\text{cyl}[\boldsymbol{Y},u]=0\,$.}
\begin{equation}
    (\partial_\sigma\boldsymbol{Y}+E\boldsymbol{Y})\big|_{\sigma=0}=0.\label{eq:Y_Robin}
\end{equation}
The relevant Green's function is hence the one satisfying $\Box G(x,x')\sim\delta(x-x')$ in the bulk, and satisfying $(\partial_\sigma +E)G(x,x')=0$ for $x$ on the boundary.

The large-$N$ analysis begins with {assuming} that the fluctuations of $u(\tau)$ are suppressed as $N\to\infty$ (so effectively it is $N^{\#>0}u(\tau)=:\psi(\tau)$ that is a well-behaved ``$\mathcal{O}(N^0)$ quantum field'' as $N\to\infty$). Then the term proportional to $u(\tau)$ in \eqref{eq:actionFinalY} can be treated as a perturbation. The justification comes after the analysis sheds light on the fluctuations of $u(\tau)$ through the saddle-point method. See e.g. Section~9.8 in \cite{Shifman:2012zz}, or Chapter 8 in \cite{polyakov2018gauge}.

The present problem admits a saddle-point solution if we take 
\begin{equation}
    \frac{1}{g_0}\propto H^2\propto N.
\end{equation}
The partition function is
\begin{equation}
\begin{split}
    Z \sim \int D\boldsymbol{Y}\! \int Du \;
    & \exp\Bigl\{ \frac{1}{4\pi} \int_\text{bulk}  \boldsymbol{Y}\cdot \Box \boldsymbol{Y} + \frac{1}{4\pi} \int_{\partial}{\mathrm d\tau \; \Bigl [ - \boldsymbol{Y}\cdot\partial_{\sigma}\boldsymbol{Y} - E \boldsymbol{Y}^2} + \frac{E}{g_0} + \frac{ (4\pi H)^2 }{E} \Bigr] \Bigr\} \\
    & \times \quad \exp \Bigl\{ -\frac{1}{4\pi} \int_{\partial}{\mathrm d\tau \; u(\tau) \Bigl[\big(\boldsymbol{Y} + \frac{4\pi \boldsymbol{H}}{E}\big)^2-\frac{1}{g_0}\Bigr]} \Bigr\}\,,
\end{split}
\end{equation}
with $\sim$ implying that the crossed-channel Casimir energy contribution $e^{\frac{N\ell}{12R}}$ and the normalization constant $N$ in \eqref{eq:ZRH_def} are suppressed for brevity.

We next perform the path integral over $\boldsymbol{Y}$. This is possible because the action is quadratic. The presence of the boundary and the boundary terms makes the calculation slightly subtle, but the following trick solves the problem. In the first step we integrate over $\boldsymbol{Y}$ subject to the boundary condition $\boldsymbol{Y}|_\partial=\boldsymbol{Y}_{\!\!B}(\tau)$ (with $\boldsymbol{Y}_{\!\!B}(\tau)$ treated as a non-fluctuating classical field configuration), and in the second step we perform the (quantum mechanical) path integral over $\boldsymbol{Y}_{\!\!B}(\tau)$. The first step is streamlined through the following standard decomposition:
\begin{equation}
    \boldsymbol{Y}=\boldsymbol{Y}_\text{\!\!cl}[\boldsymbol{Y}_{\!\!B}]+\tilde{\boldsymbol{Y}},\qquad\text{with}\qquad \tilde{\boldsymbol{Y}}\big|_\partial=0.
\end{equation}
Here $\boldsymbol{Y}_\text{\!\!cl}[\boldsymbol{Y}_{\!\!B}]$ is the classical part of $\boldsymbol{Y}$ (hence subject to $\Box \boldsymbol{Y}_\text{\!\!cl}=0$ in the bulk) ensuring that $\boldsymbol{Y}$ approaches on the boundary the (so far) non-fluctuating configuration $\boldsymbol{Y}_{\!\!B}(\tau)$. Only $\tilde{\boldsymbol{Y}}$ fluctuates in the first step. The path integral in this step is hence of standard Dirichlet type. Importantly, cancelations of the crossed terms (due to $\boldsymbol{Y}_\text{\!\!cl}$ being harmonic and $\tilde{\boldsymbol{Y}}$ vanishing on the boundary) implies
\begin{equation}
    \int_\text{bulk}\boldsymbol{Y}\cdot\Box\boldsymbol{Y}-\int_\partial\boldsymbol{Y}\cdot\partial_\sigma \boldsymbol{Y}=\int_\text{bulk}\tilde{\boldsymbol{Y}}\cdot\Box\tilde{\boldsymbol{Y}}-\int_\partial\boldsymbol{Y}_{\!\!B}\cdot\partial_\sigma \boldsymbol{Y}_\text{\!\!cl}.\label{eq:action_simp}
\end{equation}
The path integral over $\tilde{\boldsymbol{Y}}$ is therefore independent of $\boldsymbol{Y}_{\!\!B}(\tau)$, and the boundary terms all factor out of it. The result gives a bulk partition function and, in particular, cancels the pre-factors $e^{\frac{N\ell}{12R}}$ and $N$ in \eqref{eq:ZRH_def}. No $\boldsymbol{Y}_\text{\!\!cl}$ remains in the bulk, and on the boundary it is equal to $\boldsymbol{Y}_{\!\!B}(\tau)$. There is also $\partial_\sigma\boldsymbol{Y}_\text{\!\!cl}$ on the RHS of \eqref{eq:action_simp}, which may appear at first to obstruct formulating the problem as a quantum mechanical path integral (without reference to the bulk, and in particular $\boldsymbol{Y}_\text{\!\!cl}$). But that can be dealt with using the Dirichlet-to-Neumann operator $K$ defined through
\begin{equation}
    \partial_\sigma\boldsymbol{Y}_\text{\!\!cl}(\sigma,\tau)\Big|_{\sigma=0}=\int\mathrm{d}\tau'\ K(\tau-\tau')\,\boldsymbol{Y}_{\!\! B}(\tau')\,.\label{eq:K_def}
\end{equation}
We thus end up with a quantum mechanical path integral over $\boldsymbol{Y}_{\!\!B}(\tau)$ as
\begin{equation}
\begin{split}
    \int \!\!D\boldsymbol{Y}_{\!\! B}\! \int\!\! Du \,
    & \exp\Bigl\{ \int\frac{\mathrm d\tau}{4\pi} \; \Bigl [-\Big(\int \mathrm{d}\tau'\,\boldsymbol{Y}_{\!\!B}(\tau)\big(K(\tau-\tau')+E\,\delta(\tau-\tau')\big)\cdot\boldsymbol{Y}_{\!\!B}(\tau') \Big)+ \frac{E}{g_0} + \frac{ (4\pi H)^2 }{E} \Bigr] \Bigr\} \\
    & \times \quad \exp \Bigl\{ - \int{\frac{\mathrm d\tau}{4\pi} \; u(\tau) \Bigl[\big(\boldsymbol{Y}_{\!\!B}(\tau) + \frac{4\pi \boldsymbol{H}}{E}\big)^2-\frac{1}{g_0}\Bigr]} \Bigr\}\,,\label{eq:Seff_YB}
\end{split}
\end{equation}
Treating the second line as a perturbation term, we Taylor expand it and perform the Gaussian $D\boldsymbol{Y}_{\!\! B}$ path integral. Then we use\footnote{Fourier expanding along $\tau$ turns the bulk equation of motion into $(\partial_\sigma^2-\frac{n^2}{R^2})\hat{\boldsymbol{Y}}_\text{\!\!cl}(\sigma,n)=0,$ whose decaying solution as $\sigma\to-\infty$ is $\hat{\boldsymbol{Y}}_{\!\!B}(n)e^{\frac{|n|}{R}\sigma}$. This, combined with \eqref{eq:K_def}, implies $\hat K(n)=\frac{|n|}{2\pi R^2}$. Eq.~\eqref{eq:K_vs_G} is hence equivalent to $\hat G_B(n)=\frac{1}{|n|+ER}$, which is demonstrated in \eqref{eq:disc_lead_greens_func}.}
\begin{equation}
    K(\tau-\tau')+E\,\delta(\tau-\tau')=\frac{(G_B^{-1})(\tau-\tau')}{2\pi R^2},\label{eq:K_vs_G}
\end{equation}
with $G_B$ the boundary-to-boundary propagator of a single component $Y^i$ with Robin boundary condition \eqref{eq:Y_Robin}. Re-summing the Taylor expansion, we arrive at:
\begin{equation}
\begin{split}
    Z_B=\int Du \; 
    &\exp \left\{\frac{N}{2} \; \mathrm{tr} \left(\log(G_B)\right)  + \frac{1}{4\pi} \int_{\partial} \mathrm d\tau \; \left[ \frac{E}{g_0} + \frac{ (4\pi H)^2 }{E}\right] \right\} \\
    \times \quad
    &\big\langle \,\exp \left\{ -\frac{1}{4\pi} \int_{\partial} \mathrm d\tau \; u(\tau) \;\Big[  \big(\boldsymbol{Y}_B + \frac{4\pi \boldsymbol{H}}{E}\big)^2-\frac{1}{g_0}\Big]  \right\}\,\big\rangle^{}_E\,.
    \label{eq:Z_with_Y_out}
\end{split}
\end{equation}

The saddle-point equation (or the Euler-Lagrange equation for $u(\tau)$) reads:
\begin{equation}
    \frac{1}{g_0} = \langle \boldsymbol{Y}_B^2 \rangle^{}_E + \frac{16\pi^2\boldsymbol{H}^2}{E^2}.
    \label{eq:disc_saddle_point_eq}
\end{equation}
Note that the expectation values $\langle\cdot\rangle^{}_E$ are with respect to $\boldsymbol{Y}_{\!\! B}$ as the random variable.

To evaluate $\langle \boldsymbol{Y}_B^2 \rangle^{}_E$, we use the well-known trick of connecting it with the $\boldsymbol{Y}$-field's Green's function (see e.g. \cite{Witten:1992}).
For a single component of $\boldsymbol{Y}$, the Green's function, described below \eqref{eq:Y_Robin}, is given on the disk explicitly by:
\begin{equation}
\begin{split}
        G(z, w) = -\log\frac{|z-w|}{R}-\log\Big|1-\frac{z\bar{w}}{R^2}\Big| + \frac{1}{ER} - \sum_{n=1}^\infty {\frac{ER}{n(n+ER)}\Big[\big(\frac{\bar{z}w}{R^2}\big)^n + \big(\frac{z\bar{w}}{R^2}\big)^n\Big]}\,,
    \label{eq:greens_func_lead}
\end{split}
\end{equation}
where $z=re^{i\theta}$ and $w=r_0e^{i\theta_0}$ are the observation and source points, respectively. On the boundary $r=R$, we get:
\begin{equation}
    G_B(\theta, \theta_0) = \sum_{n=-\infty}^\infty {\frac{e^{in(\theta-\theta_0)}}{|n|+ER}}
    \label{eq:disc_lead_greens_func}
\end{equation}
The coincidence limit of the boundary-to-boundary 2-point function in the large-$R$ limit is:
\begin{equation}
    \langle \boldsymbol{Y}_B^2 \rangle^{}_E = N \; \lim_{\theta \rightarrow \theta_0} G_B(\theta, \theta_0) = -2N \Bigl[\log(ER) +\gamma + \log(\epsilon) \Bigr],
    \label{eq:coincidental_two_point_def}
\end{equation}
with $\epsilon$ a short-distance cutoff on the boundary. We can now write the saddle-point equation \eqref{eq:disc_saddle_point_eq} as:
\begin{equation}
    -2N \Bigl[\log(ER) +\gamma + \log(\epsilon) \Bigr] + \frac{16 \pi^2 \boldsymbol{H}^2}{E^2} = \frac{1}{g_0}.
    \label{eq:saddle-point-epsilon}
\end{equation}

To renormalize our leading-order results, let us define an energy scale $E_0$ as the value of $E$ at $H=0$:
\begin{equation}
    -2N\big[\log(E_0R)+\gamma + \log(\epsilon)\big] = \frac{1}{g_0},
    \label{eq:E0_vs_g0}
\end{equation}
We will shortly relate $E_0$ to the physical scale $E^\ast$ using the normalization condition \eqref{eq:norm_cond_intro}. Equations~\eqref{eq:saddle-point-epsilon} and \eqref{eq:E0_vs_g0}, on the other hand, allow expressing $E$ in terms of $E_0$ as\footnote{To obtain $E(H,E_0)$ from $E^2\log\frac{E}{E_0}=\frac{8\pi^2H^2}{N},$ note that $y^2\log y=t$ can be inverted to $y(t)=\sum_{k=0}^\infty\frac{(1-2k)^{k-1}}{k!}t^k$ by writing $y=e^z,$ $t=z\,e^{2z},$ and then evaluating $y_k=\oint\frac{y(t)}{t^{k+1}}\frac{\mathrm{d}t}{2\pi i}=\frac{1}{k}\oint\frac{e^{-(2k-1)z}}{z^k}\frac{\mathrm{d}z}{2\pi i }$. Alternatively, setting $W(x)=\frac{1}{2}\log y$, one can recognize \cite{Corless1996LambertW} the Lambert W function $W(x)\exp{W(x)}=x:=2t,$ that upon application of the Lagrange inversion formula \cite{FlajoletSedgewick2009} $W(x)^r = \sum_{n=r}^{\infty} -\frac{r(-n)^{n-r-1}}{(n-r)!}x^n,
$ in the special case $r=-1/2$ produces the series for $y(t).$}
\begin{equation}
    \frac{E}{E_0} = 1 + \sum_{k=1}^\infty {\frac{(1-2k)^{k-1}}{2^k k!} \left[ \frac{1}{N} \left( \frac{4\pi H}{E_0} \right)^2 \right]^{k}}.
    \label{eq:E_exp}
\end{equation}

Recall that our aim is to calculate the leading boundary free energy, which according to \eqref{eq:Z_with_Y_out}, is given by:
\begin{equation}
    F= -\log(Z_B) = -\frac{N}{2} \mathrm{tr}\big(\log(G_B)\big) - \frac{ER}{2} \left[ \frac{1}{g_0} + \left( \frac{4\pi H}{E} \right)^2 \right].
    \label{eq:disc_free_energy_def}
\end{equation}
Using \eqref{eq:disc_lead_greens_func} and \eqref{eq:coincidental_two_point_def}, the divergent piece $\mathrm{tr}\big(\log(G_B)\big) = \lim_{\theta \rightarrow \theta_0} \sum^{\infty}_{n=-\infty} {\log\Bigl( \frac{e^{in(\theta-\theta_0)}}{|n|+ER}} \Bigr)$ has the following relation with $\langle \boldsymbol{Y}_B^2 \rangle^{}_E$:
\begin{equation}
   N \frac{\mathrm d}{\mathrm d(ER)} \mathrm{tr}\big(\log(G_B)\big) = -\langle \boldsymbol{Y}_B^2 \rangle^{}_E = -\frac{1}{g_0} + 2N \log(\frac{E}{E_0}),
    \label{eq:coincidental_vs_derivative}
\end{equation}
with the last equality following from the saddle-point equation \eqref{eq:disc_saddle_point_eq}. Integrating \eqref{eq:coincidental_vs_derivative} back and substituting in \eqref{eq:disc_free_energy_def}, we get the renormalized free energy:
\begin{equation}
\begin{split}
        F = NER \Bigl[1-2\log\Big(\frac{E}{E_0}\Big)\Bigr] + \frac{C}{2}\,,
    \label{eq:leading_F}
\end{split}
\end{equation}
where $C$ is the integration constant. We have found that an alternative Pauli-Villars regularization of $\mathrm{tr}\big(\log G_B(0)\big)$ yields $C=0.$ We therefore discard $C$ in \eqref{eq:leading_F}, and expand $\mathcal{E}=\frac{F}{2\pi R}$ using \eqref{eq:E_exp} to arrive at:
\begin{equation}
    \boxed{\mathcal{E}_{\text{lead}} =  \frac{E_0}{4\pi}\Bigl[2N - \Big(\frac{4\pi H}{E_0}\Big)^2 + \frac{1}{4N} \Big(\frac{4\pi H}{E_0}\Big)^4 - \frac{5}{24N^2} \Big(\frac{4\pi H}{E_0}\Big)^6 + ...\Bigr]}.
    \label{eq:leading_fe}
\end{equation}
We can now use the normalization condition (\ref{eq:norm_cond_intro}) to obtain
\begin{equation}
    \boxed{\frac{1}{E^*} = -\frac{1}{2} \frac{\partial^2 \mathcal{E}}{\partial H^2}\big|^{}_{H=0}\, \overset{\text{lead}}{\Longrightarrow}E_0=4\pi E^\ast.}
    \label{eq:norm_cond_lead}
\end{equation}
It is straightforward to check that with this $E_0$, Eq.~\eqref{eq:leading_fe} verifies the Lukyanov-Zamolodchikov conjecture \eqref{eq:sasha_full_coefs} at the leading order in the $1/N$ expansion.

\section{Next-to-leading order solution in the $1/N$ expansion}\label{sec:subleading}

The exponential on the second line of \eqref{eq:Z_with_Y_out} can be expanded as a perturbation. We keep only the connected terms and sum back as an exponential. This yields:
\begin{equation}
\begin{split}
    & \frac{Z}{Z_\text{lead}} = \int Du \; e^{W[u]} \\
    & W[u] = -\frac{1}{4\pi} \int_{\partial}{\mathrm d\tau \; u(\tau) \Bigl[\langle\boldsymbol{Y}^2(\tau)\rangle_\text{conn.} +\, \Bigr(\frac{4\pi H}{E}\Bigr)^2-\frac{1}{g_0}\Bigr]} \\
    & + 2 \int_{\partial} \int_{\partial} \mathrm d\tau\, \mathrm d\tau' \; u(\tau)\, u(\tau') \; \frac{1}{E^2}\; \langle\boldsymbol{Y}(\tau)\cdot\boldsymbol{H} \; \boldsymbol{Y}(\tau')\cdot\boldsymbol{H}\rangle_\text{conn.} \\
    & + \frac{1}{2!} \frac{1}{(4\pi)^2} \int_{\partial} \int_{\partial} \mathrm d\tau \,\mathrm d\tau' \; u(\tau)\, u(\tau') \langle \boldsymbol{Y}^2(\tau)\boldsymbol{Y}^2(\tau')\rangle_\text{conn.} \\
    & + \quad ...\label{eq:sub_effective_action}
\end{split}
\end{equation}

The first term on the RHS of $W[u]$ vanishes by definition in the saddle-point approximation, see \eqref{eq:disc_saddle_point_eq}. Using the leading $Y$-field propagator as in \eqref{eq:disc_lead_greens_func}, and writing 
\begin{equation}
    u(\tau) = \sum_{n_1=-\infty}^\infty {\frac{\tilde{u}_{n_1}}{R}\,e^{\frac{in_1\tau}{R}}},
\end{equation}
we can evaluate the second and third terms with the following Feynman graph representations in Fourier space:
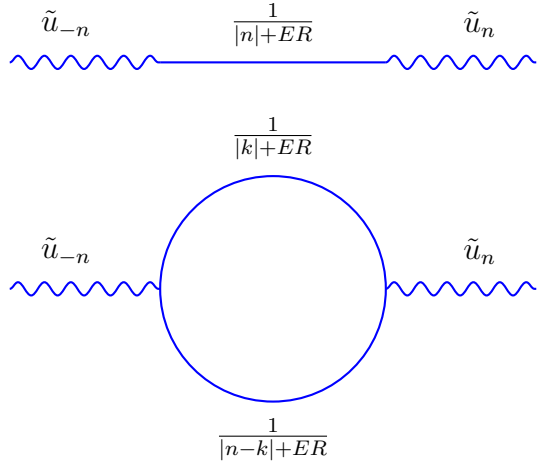
\begin{center}
\begin{tikzpicture}[thick]
  \path [draw=blue,snake it]
    (-3,0) -- (-1,0);
  \path [draw=blue]
    (-1,0) -- (2,0);
  \path [draw=blue,snake it]
    (2,0) -- (4,0);
  \node[align=left] at (-2.25,0.5) {$\tilde{u}_{-n}$};
  \node[align=left] at (3.25,0.5) {$\tilde{u}_n$};
  \node[align=center] at (0.5,0.5) {$\frac{1}{|n|+ER}$};
\end{tikzpicture}
\end{center}

\begin{center}
\begin{tikzpicture}[thick]
  \path [draw=blue,snake it]
    (-3,0) -- (-1,0);
  \draw[draw=blue] (2,0) arc (0:360:1.5cm);
  \path [draw=blue,snake it]
    (2,0) -- (4,0);
  \node[align=left] at (-2.25,0.5) {$\tilde{u}_{-n}$};
  \node[align=left] at (3.25,0.5) {$\tilde{u}_n$};
  \node[align=center] at (0.5,2) {$\frac{1}{|k|+ER}$};
  \node[align=center] at (0.5,-2) {$\frac{1}{|n-k|+ER}$};
\end{tikzpicture}
\end{center}

The result is
\begin{equation}
    \frac{Z_\text{lead+sub}}{Z_\text{lead}} = \int Du \; \exp\Bigl[-\sum_{n=-\infty}^\infty \tilde{u}_{-n} \; K(n) \; \tilde{u}_{n} \Bigr]\,,
    \label{eq:z-by-z-lead}
\end{equation}
with
\begin{equation}
    K(n) = \frac{N}{2} \; \Bigl[ {\frac{L}{|n|+ER}} + \frac{1}{2} \sum_{k=-\infty}^\infty {\frac{1}{|k| + ER} \frac{1}{|n-k| + ER} } \Bigr]\,.
    \label{eq:sub_propagator_raw}
\end{equation}
Here, we used the short-hand
\begin{equation}
    L \equiv \frac{1}{N}\Bigl(\frac{4\pi H}{E}\Bigr)^2.
\end{equation}
The inner sum has a closed-form expression in terms of the digamma function $\psi$:
\begin{equation}
    \frac{1}{2} \sum_{k=-\infty}^\infty {\frac{1}{|k| + ER} \frac{1}{|n-k| + ER} }= \frac{1}{ER \, n}+2\frac{n+ER}{n(n+2ER)}\left[\psi(ER+n)-\psi(ER+1)\right].
    \label{eq:inner-sum-closed-form}
\end{equation}

We take the Lagrange multiplier field, $u(\tau)$, to be real, so that $\tilde{u}_{-n} = \tilde{u}^*_{n}$. Then using $\tilde{u}_n = \xi_n + i \; \eta_n$, with real $\xi_n$ and $\eta_n$, we have:
\begin{equation}
\begin{split}
    \frac{Z_\text{lead+sub}}{Z_\text{lead}}
    & = \int Du \; \exp\Bigl[-\sum_{n=-\infty}^\infty |\tilde{u}_n|^2 \; K(n) \Bigr] = \int \mathrm d\xi_0\ \mathcal{N}\; e^{-K(0) \xi_0^2} \; \prod_{n=1}^{\infty} \int \mathrm d\xi_n \,\mathrm d\eta_n\  \mathcal{N}^2 \; e^{-K(n) (\xi_n^2 + \eta_n^2)} \\
    & = \prod_{n=-\infty}^{\infty} \left[\frac{\pi \mathcal{N}^2}{K(n)} \right]^{\frac{1}{2}} = \exp\left[-\frac{1}{2} \sum_{n=-\infty}^\infty \log(\frac{K(n)}{\pi \mathcal{N}^2})\right]
    \label{eq:sub_partition_func_raw}
\end{split}
\end{equation}

The normalization factor $\mathcal{N}$ is required when we discretize the path integral. This is evident already in the calculation of the quantum mechanical path integral:
\begin{equation}
    [Dq(\tau)] = \lim_{n \rightarrow\infty} \left(\frac{1}{2\pi \Delta \tau} \right)^{\frac{n}{2}} \prod_{k=1}^{n-1} dq_k
    \label{eq:qm_normalization}
\end{equation}
Importantly, as $n\to\infty$, if the time direction is compactified with length $l$, it is not $\Delta\tau$ that is kept fixed as $n\to\infty$ but rather $n\Delta\tau=l.$ We would thus get a factor $\propto l^{-\frac{n}{2}}.$

From \eqref{eq:sub_partition_func_raw}, we obtain the subleading part of the free energy, $F_\text{lead+sub} - F_\text{lead}=-\log(\frac{Z_\text{lead+sub}}{Z_\text{lead}})$, ignoring the irrelevant constants, to be
\begin{equation}
    F_\text{sub}=\frac{1}{2}\sum_{n=-\infty}^\infty \log{\frac{K(n)}{\mathcal{N}^2}}\,.
\end{equation}
In the large-$R$ limit of our interest, we can replace the sum with an integral and the digamma function with its large-argument asymptotics (see Appendix~\ref{app:justif}), giving
\begin{equation}
    \mathcal{E}_{\text{sub}}=\frac{F_{\text{sub}}}{2\pi R} \xrightarrow{R\to\infty} \frac{1}{2\pi}\int_{0}^\Lambda \mathrm dp \; \log\Big(\frac{N}{2R\mathcal{N}^2}\bigl[\frac{L}{p+E} + 2\frac{p+E}{p(p+2E)} \log\big(1+\frac{p}{E}\big) \bigl]\Big)
    \label{eq:fe_2}
\end{equation}

\subsection{Renormalizing the subleading solution}

Taylor expanding \eqref{eq:fe_2} in powers of $L_0\equiv \frac{1}{N} \Big(\frac{4\pi H}{E_0}\Big)^2$ yields:
\begin{equation}
    \boxed{\mathcal{E}_{\text{sub}} =I_0+\sum_{k=1}^\infty I_k \frac{1}{N^k} \Big(\frac{4\pi H}{E_0}\Big)^{2k}\,,}
\end{equation}
where $I_k$ are UV-regulated integrals. For example
\begin{equation}
\begin{split}
    I_1 =\frac{E_0}{4\pi} \Big[\gamma+\log\Big(\frac{4E_0}{\Lambda} \; \log(\frac{\Lambda}{E_0})\Big) \Big].
    \label{eq:I1}
\end{split}
\end{equation}
The logarithmic UV divergence in $I_1$ is ``\emph{primitive}'', in the sense that the UV divergence in all $k \ge 1$ terms is a multiple of $I_1$: 
\begin{equation}
    I_k=\frac{E_0}{4\pi} \frac{(1-2k)^{k-2}}{k! \; 2^{k-1}} \Big\{k-1+(1-2k) \big[\log(2k-1)+\frac{4\pi I_1}{E_0} \big]\Big \}\,.
    \label{eq:sub_integrals}
\end{equation}
All these UV divergences can hence be renormalized away with a single normalization condition, namely \eqref{eq:norm_cond_intro}.

At the leading level, from the normalization condition \eqref{eq:norm_cond_intro} we deduced that $E_0=4\pi E^\ast$. At the subleading level, this relation receives a correction
\begin{equation}
    \boxed{\frac{1}{E^*} = -\frac{1}{2} \frac{\partial^2 \mathcal{E}}{\partial H^2}|_{H=0} \overset{\text{lead+sub}}{\Longrightarrow} \frac{1}{E^*} 
    = \frac{4\pi}{E_0} \Big(1-\frac{4\pi I_1}{NE_0}\Big)}\,.
    \label{eq:sub_ct}
\end{equation}
Substituting the above modified relation in the leading boundary energy \eqref{eq:leading_fe} produces a counter-term at the subleading level which removes the divergence from all the $k \ge 1$ integrals \eqref{eq:sub_integrals}. It is straightforward to check that the resulting finite answer verifies the Lukyanov-Zamolodchikov conjecture \eqref{eq:sasha_full_coefs} at the next-to-leading order in the $1/N$ expansion.

The only remaining UV divergence is that of $I_0$, which takes the form
\begin{equation}
    I_0=-\frac{E_0}{2\pi} +\left[ -\frac{E_0}{2\pi} \,\mathrm{li}(1+\frac{\Lambda}{E_0}) + \frac{\Lambda}{2\pi} + \frac{\Lambda}{2\pi} \log \big(\log(\frac{\Lambda}{E_0})\big) +\frac{\Lambda}{2\pi} \log\big(\frac{N}{R \mathcal{N}^2 \Lambda}\big)  \right].\label{eq:I0_simp}
\end{equation}
To remove the UV-divergent terms in the square brackets, we first note that the factor $\mathcal{N}^2$ (analogous to the factor in \eqref{eq:qm_normalization}), must cancel the volume $2 \pi R$, so that $\mathcal{N}^2 R$ is independent of $R$. This is expected on general grounds since UV divergences should not depend on IR data such as $R\,.$ (Our analysis of the half-plane solution in the next subsection clarifies this point.) The whole piece in the square brackets in \eqref{eq:I0_simp} can then be removed by a boundary cosmological constant counterterm.

\subsection{Comparison with the half-plane}

The effective action for the half-plane domain is the same as \eqref{eq:sub_effective_action}, except that $\int_0^{2\pi R} \mathrm{d}\tau$ on the boundary $\sigma=0$ is replaced with $\int_{-\infty}^{\infty} \mathrm{d}x$ on the boundary $y=0$. Since the domain is non-compact, instead of \eqref{eq:disc_lead_greens_func} the Green's function is 
\begin{equation}
    G_B(x,x_0) = \int \mathrm{d}p \; \frac{e^{ip(x-x_0)}}{|p|+E}.
\end{equation} 

Following similar steps as in the previous section, we get
\begin{equation}
    \frac{Z_\text{lead+sub}}{Z_\text{lead}} = \int {D}u \; e^{-\int_{-\infty}^\infty \mathrm{d}p \; \tilde{u}(-p) \; K(p) \; \tilde{u}(p)},
    \label{eq:hp-z-by-z-lead}
\end{equation}
with
\begin{equation}
    K(p) = \frac{N}{2} \; \Bigl[ {\frac{L}{|p|+E}} + \frac{1}{2} \int_{-\infty}^\infty {\frac{1}{|k| + E} \frac{1}{|p-k| + E} } \Bigr]\,.
    \label{eq:sub_propagator_raw}
\end{equation}

We calculate the partition function \eqref{eq:hp-z-by-z-lead} in the Fourier space, by converting the measure into an infinite product of $d\Im{\tilde{u}(p)} \; d\Re{\tilde{u}(p)}$ at all $p$. Since $\tilde{u}(p)$ and $\tilde{u}(-p)$ are not independent, we only consider $p \ge 0$:
\begin{equation}
\begin{split}
    \frac{Z_\text{lead+sub}}{Z_\text{lead}} = \prod_{p \ge 0} \left[ \left(\mathcal{N}\, \mathrm{d}\Im{\tilde{u}(p)} \right) \; \left(\mathcal{N} \, \mathrm{d}\Re{\tilde{u}(p)} \right) \right] \; \exp\left[\int_0^{\infty} \mathrm{d}p \; \left(\Im{\tilde{u}^2(p)} + \Re{\tilde{u}^2(p)} \right) K(p) \right]\,.
    \label{eq:hp_partition_1}
\end{split}
\end{equation}

We discretize the exponent in \eqref{eq:hp_partition_1} by assuming a finite length $l$ for the domain and then later sending $l$ to $\infty$:
\begin{equation}
    \int_0^{\infty} \mathrm{d}p \; |\tilde{u}(p)|^2 K(p) = \frac{2\pi}{l}\sum_{p=0}^{\infty} (\Im{\tilde{u}^2(p)} + \Re{\tilde{u}^2(p)}) K(p)
\end{equation}
\begin{equation}
\begin{split}
    \Longrightarrow\frac{Z_\text{lead+sub}}{Z_\text{lead}} &= \mathcal{N}\, \int \mathrm{d}\Im{\tilde{u}(0)} \; e^{-\frac{2\pi}{l} \Im{\tilde{u}^2(0)} \, K(0)} \\
    &\ \ \ \times \prod_{p > 0} \left[ \mathcal{N}\, \int \mathrm{d}\Im{\tilde{u}(p)} \; e^{-\frac{2\pi}{l} \Im{\tilde{u}^2(p)} \, K(p)} \right] \\
    &\ \ \ \times \prod_{p > 0} \left[ \mathcal{N} \, \int \mathrm{d}\Re{\tilde{u}(p)} \; e^{-\frac{2\pi}{l} \Re{\tilde{u}^2(p)} \, K(p)} \right]\,.
    \label{eq:hp_partition_2}
\end{split}
\end{equation}
The integrals are all identical ordinary Gaussians and the result is:
\begin{equation}
    \frac{Z_\text{lead+sub}}{Z_\text{lead}} = \prod_{p=-\infty}^{\infty} \mathcal{N} \sqrt{\frac{l}{2K(|p|)}}\,.
\end{equation}

The subleading free energy is obtained by finding the negative logarithm of the partition function, that is:
\begin{equation}
    F_\text{sub} = \frac{1}{2} \sum_{p=-\infty}^{\infty} {\log{\frac{2K(|p|)}{\mathcal{N}^2 \, l}}}\,.
\end{equation}
We substitute for $K(|p|)$ using by integrating for $p \ge 0$:
\begin{equation}
    \int_{-\infty}^\infty {\frac{1}{|k| + E} \frac{1}{|p-k| + E} } = \frac{4(E + p)}{p(2E + p)} \log\left(1 + \frac{p}{E} \right).
    \label{eq:hp_u_propagator_integrated}
\end{equation}
Then sending $l$ to $\infty$, the sum becomes an integral:
\begin{equation}
    \mathcal{E}_\text{sub} = \frac{F_\text{sub}}{l} \xrightarrow{l\to\infty} \frac{1}{2\pi} \int_{0}^{\Lambda} \mathrm{d}p \; {\log { \left\{ \frac{N}{\mathcal{N}^2 \; l}\, \left[ \frac{L}{p+E}+2\frac{E + p}{p(2E + p)} \log \left(1 + \frac{p}{E} \right)  \right] \right\} }}\,.
\end{equation}

As expected, this integral is the same as \eqref{eq:fe_2} of the disk domain. Similarly to the disk case, we argue that the factor $\mathcal{N}^2$ cancels the infinite volume $l$, so that $\mathcal{N}^2\,l$ is independent of $l$. This is usual in evaluating path integrals by discretization; see \eqref{eq:qm_normalization}.

\subsection{Connection with the RG-running of $g_0$}

Taking logarithm of the normalization condition at the subleading order, Eq.~\eqref{eq:sub_ct}, expanding $\log(1-\frac{4\pi I_1}{NE_0})$ in small $\frac{4\pi I_1}{NE_0}$, and using the definition of $E_0$ in Eq.~\eqref{eq:E0_vs_g0}, the $1/N$ correction in Eq.~\eqref{eq:sub_ct} is equivalent to the following identity:
\begin{equation}
    \log(\frac{\Lambda}{4\pi E^*}) - \frac{1}{2Ng_0} \approx -\frac{1}{N} \; \left[\gamma + \log(4) -\frac{1}{2Ng_0}-\log(2Ng_0)\right].
\end{equation}
Taking $\Lambda \partial_\Lambda$ converts this to:
\begin{equation}
    \Lambda \partial_\Lambda \frac{1}{g_0} \approx 2N[1+\frac{1}{N}+2g_0]^{-1} \approx 2(N-1)+4Ng_0\,.
    \label{eq:beta_from_ct}
\end{equation}
This matches the Gellmann-Low RG equation obtained in \cite{Giombi:2019enr}:
\begin{equation}
    \Lambda \partial_\Lambda \frac{1}{g_0} = 2(N-1) (1+2g_0)+...,
    \label{eq:giombi_beta}
\end{equation}
up to subleading order in $1/N$, using $g_0=\mathcal O(1/N)$. Therefore, our next-to-leading order calculation is consistent with the RG running of $g_0$.

\section{Future directions}\label{sec:future}

In this section we outline some directions for future research, and in particular some avenues for extending our results beyond verification of the Lukyanov-Zamolodchikov conjecture \eqref{eq:sasha_full_coefs}.

\subsection{Extension to $C_0$}

Our treatment of the boundary free energy, while sufficient for verifying the conjecture \eqref{eq:sasha_full_coefs} for $C_{k\ge1}$, leaves some ambiguities in the choice of a normalization condition relevant to $C_0$. 

At the leading order, we regularized the boundary free energy by connecting it with the saddle-point equation, as in \eqref{eq:coincidental_vs_derivative}. The resulting integration required fixing an additive constant, which we determined through an alternative Pauli–Villars calculation. The ensuing Eq.~\eqref{eq:leading_fe} then gives $C_0$ to be $2N$. In \cite{Lukyanov:2003rt}, the authors suggested that a natural value for $C_0$ would be the real part of the analytic continuation of $C_{k\ge1}$ to $k=0$:
\begin{equation}
    C_0\overset{?}{=}2\pi \cot\big(\frac{\pi}{N-1}\big)=2N-2+\mathcal{O}(1/N),\label{eq:C0?}
\end{equation}
which in the large-$N$ limit indeed matches the result $C_0\approx 2N$ found (within the Pauli-Villars regularization scheme) in Eq.~\eqref{eq:leading_fe}.

At the subleading order, we further simplified \eqref{eq:I0_simp} by discarding the UV-divergent piece inside the square brackets. In the regularization scheme set by that particular counterterm, we get the first term $-\frac{E_0}{2\pi}$ in Eq.~\eqref{eq:I0_simp} for $I_0\,.$ This indeed reproduces the subleading piece of \eqref{eq:C0?}.



A more systematic approach would be to formulate an explicit \emph{normalization condition}. In elementary quantum field theory, the number of such conditions is typically taken to match the number of \emph{primitive divergences}---those irreducible divergences in Feynman diagrams from which all others can be derived. In our setting, one such normalization condition, following \cite{Lukyanov:2003rt}, was imposed on $E^\ast$ in \eqref{eq:norm_cond_intro}. The corresponding primitive divergence appeared in $I_1$ as shown in \eqref{eq:I1}--\eqref{eq:sub_integrals}. However, $I_0$ contains an additional primitive UV divergence that cannot be reduced to that of $I_1$. This observation suggests that a second normalization condition should naturally accompany our analysis. The new normalization condition would fix the boundary cosmological constant counterterm, including the finite piece which would contribute to $C_0.$


As usual, the choice of a normalization condition is not unique. One can simply choose \eqref{eq:C0?} as the normalization condition, or even $C_0=0$. Alternatively, in the context of the ODE/IM correspondence, one may choose the normalization condition fixing $C_0$ to the result on the ODE side.\footnote{The ODE result is not currently available at large $N.$ However, as a proof of principle one may consider $N=3$ where it should be possible to use the spherical limit of the pillow-brane result \cite{Lukyanov:2012wq} to set a normalization condition for $C_0$ on the field theory side. We thank S.~Lukyanov for correspondence on this point.}

\subsection{Extension to finite $R$}

Our subleading order calculations were performed in the $R\to\infty$ limit in this work. It would be nice to extend our analysis to finite $R$. This would allow, among other things, computation of the boundary entropy and verification of the $g$ theorem \cite{Affleck:1991tk} as well as the Friedan-Konechny gradient formula \cite{Friedan:2003yc}.

\begin{acknowledgments}
    We are indebted to A.~Zamolodchikov for introducing us to this problem and for guiding us throughout the project. We also thank S.~Lukyanov, F.~Popov, A.~Raviv-Moshe, and M.~Ro\v{c}ek for related conversations. R.M.Y. also thanks A.~Bedroya, S.H.~Fadda and M.~Litvinov for illuminating conversations in the early stages of this work.
\end{acknowledgments}

\appendix

\section{Some asymptotic analysis}\label{app:justif}

\subsection{Euler-Maclaurin justification of the sum-to-integral approximation} 

In this section, we check that converting the series expression for the free energy, $\frac{1}{2}\sum_{n=-\Lambda_n}^{\Lambda_n}\log K(n)=\sum_{n=0}^{\Lambda_n}\log K(n)-\frac{\log K(0)}{2}$, to an integral as in \eqref{eq:fe_2} is justified. 

Define
\begin{equation}
\begin{split}
    f&
    (n):=\log K(n)-\log N\\
    &=\log\Big(\frac{L}{2(n+ER)}+\frac{1}{2ERn}+\frac{n+ER}{n(n+2ER)}[\psi(ER+n)-\psi(ER+1)]\Big).
\end{split}
\end{equation}
We now use the Euler-MacLaurin formula for the difference between the integral and the sum up to the second order:
\begin{equation}
    \sum_{n=a}^b {f(i)} - \int_a^b {f(x) \; \mathrm dx} = \frac{f(a)+f(b)}{2}+\frac{f'(b)-f'(a)}{6\cdot 2!}-R_2,
    \label{eq:EM}
\end{equation}
in which
\begin{equation}
    |R_2| \le \frac{1}{12} \int_a^b {|f''(x)| \; \mathrm dx}\,.
    \label{eq:R2}
\end{equation}

We evaluate $f$ and its derivatives at $n\rightarrow 0$, as well as at $n=\Lambda_n \rightarrow \infty$, and then we find the large-$R$ asymptotics of these quantities:
\begin{equation}
    \begin{split}
    \lim_{n \rightarrow 0} f(n) &= \log \left( \frac{L}{2 ER}-\frac{1}{(2ER)^2}+\frac{\psi^1(ER)}{2} \right) \qquad\quad  f(\Lambda_n)=\log \frac{\log \Lambda_n}{\Lambda_n}+\mathcal O(1/\Lambda_n)\,,\\
    &=\log\frac{1+L}{2ER}+\mathcal O(1/(ER)^2)\\
    \lim_{n \rightarrow 0} f'(n) &=\mathcal O(1/ER),  \qquad\qquad\hspace{1.8cm} \qquad \qquad \quad f'(\Lambda_n)=\mathcal O(\frac{1}{\Lambda_n})\,.
\end{split}
\end{equation}
Since we are looking for $\mathcal{E}=\frac{\text{free energy}}{2\pi R}$ at large $R$, the $f(n)$ and $f'(n)$ terms in \eqref{eq:EM} at zero and infinity do not make a contribution. It remains to evaluate $R_2$ using \eqref{eq:R2}. 

We want to use $\int f''=f'_2-f'_1$ over the sign-definite regions of $f''$, which we will argue there are finitely many, and then argue that $f'$ at the zeros of $f''$ (or at 0 and infinity) does not have a term proportional to $R$ as $R\to\infty$. (We in fact have a stronger result: $\lim_{R\to\infty} |\int f''|=0$.) This would imply that $|\int f''|$ does not have a linear term in $R$ as $R\to \infty$, validating our sum-to-integral replacement.

To use $\int f''=f'_2-f'_1$, we first demonstrate that $f'$ is bounded. Since we are interested in small positive $L$, it is sufficient to check boundedness for $L=0$. This is easy to establish using the explicit form if one separates $n=0$ and $n>0$.\footnote{In the large-$R$ limit, in fact $f'(n)\to0$ for any $n>0$, whether $n$ is much larger or smaller than $ER$. Also $\lim_{n\to\infty} f'(n)=0$.}

Zeros of $f''$ can be located as follows. Since we are interested in small positive $L$, it is sufficient to locate the zeros for $L=0$, and then perturb them if necessary.
Define $x$ through $n= ER\cdot x$.
Then from large-$x$ asymptotics we find that there are no zeros for $n\gg ER$.
To see if there are zeros for $n$ of order $ER$, we look at the large-$ER$ asymptotics of $f''$ with $x$ of order one. This shows only a single zero at some $n^\ast$.
The remaining range $x\ll 1$, corresponding to $n$ of order one as $ER\to\infty$, is irrelevant, because we would be integrating $f''$, which is $\mathcal O(1/(ER)^2)$, over a domain that is $< \mathcal O(ER)$ in length. This would imply that we do not get a piece proportional to $R$.

The formula $\int f''=f'_2-f'_1$ can then be used for the left region where $f''$ is negative, and the right region where $f''$ is positive. The value of $f'$ at $n=0,\infty$, as well as $n^\ast$ is important then. Explicit evaluation shows that they are all $<\mathcal O(1/ER)$, so we do not get a piece proportional to $R$. 
(In fact, we get zero for all of them as $R\to\infty$.)

\subsection{Justifying the use of digamma asymptotics}

In the above discussion, we showed that the sum and the integral produce the same result in the large-$R$ regime. However, in Section~\ref{sec:subleading}, we also substituted the inner sum \eqref{eq:inner-sum-closed-form} with its asymptotic form in \eqref{eq:fe_2}:
\begin{equation}
    \frac{1}{ERp}+2\frac{p+E}{p(p+2E)} \left(\psi(R(p+E))-\psi(1+ER) \right) \ \longrightarrow \ 2\frac{p+E}{p(p+2E)}\log(1+\frac{p}{E}).
    \label{eq:psi-to-log}
\end{equation}
This is equivalent to the replacement $A(p)\to A_\text{approx}(p)$:
\begin{equation}
    A(p) \equiv \frac{p+2E}{2ER(p+E)}+ \psi(R(p+E))-\psi(1+ER) \longrightarrow A_\text{approx}(p)\equiv \log(1+\frac{p}{E}).
    \label{eq:psi-to-log-modified}
\end{equation}

Using the following known bounds on the digamma function
\begin{equation}
    \log(x+\frac{1}{2}) -\frac{1}{x} \le \psi(x) \le \log x - \frac{1}{2x} \, ,
\end{equation}
we can show that $A(p)$ in \eqref{eq:psi-to-log-modified} is bounded by
\begin{equation}
    A_\text{min}(p)=\log{\left(R(p+E)+\frac{1}{2}\right)} - C_1\,,
\end{equation}
and
\begin{equation}
    A_\text{max}(p)=\log{\left(R(p+E)\right)} - C_2,
\end{equation}
with $C_1=\psi(ER)+\frac{1}{ER}$ and $C_2=\psi(ER)+\frac{1}{2ER}$.\\

We integrate logarithm of these expressions to find the free energy:
\begin{equation}
\begin{split}
    I_\text{approx} &= \int_0^\Lambda \log (A_\text{approx}) = \int_0^\Lambda {\log \left(\log {(1+ \frac{p}{E})}\right) \, dp} \\
    &= R \, \left [ ( E+\Lambda) \log \log (1+\frac{\Lambda}{E}) -E \, \mathrm{li}(1+\frac{\Lambda}{E}) + E\gamma \right ]\,.
    \label{eq:I-approx}
\end{split}
\end{equation}

Since $A_{min}$ and $A_{max}$ have a similar form, we can integrate their $\log$ using
\begin{equation}
\begin{split}
    \int \log \left(\log(x))-C \right) \; dx = -e^C \, \mathrm{Ei} \left(-C+\log{x} \right) + x \, \log{\left(-C+\log{x}\right)}
    \label{eq:generic-integral}
\end{split}
\end{equation}

When we divide by $2\pi R$ (since we are interested in $\mathcal{E}=\frac{\text{free energy}}{2 \pi R}$) for both the upper and lower limit we get the same result as \eqref{eq:I-approx} at large $R$. This means that in the large-$R$ regime, the substitution \eqref{eq:psi-to-log} does not change the result for the specific free energy.

\bigskip

\bibliographystyle{JHEP}
\bibliography{refs}

\end{document}